\documentclass[sigconf, nonacm, manuscript]{acmart}
\AtBeginDocument{%
  \providecommand\BibTeX{{%
    \normalfont B\kern-0.5em{\scshape i\kern-0.25em b}\kern-0.8em\TeX}}}

\setcopyright{rightsretained}
\copyrightyear{2024}
\acmYear{2024}
\acmDOI{XXXXXXX.XXXXXXX}

\acmConference[XX]{XX}{XX}{XX}
\acmISBN{XX}

\begin{document}

\title[The Legend of Holy Sword]{The Legend of Holy Sword: An Immersive Experience for Concentration Enhancement}


\author{Hirosuke Asahi}
\authornote{All authors contributed equally to this research.}
\email{hirosuke.asahi@star.rcast.u-tokyo.ac.jp}
\orcid{0009-0009-7197-9248}
\affiliation{
  \institution{The University of Tokyo}
  \city{Tokyo}
  \country{Japan}
}

\author{Ryoma Sonoyama}
\orcid{0009-0008-1538-6379}
\email{ryoma.sonoyama@star.rcast.u-tokyo.ac.jp}
\affiliation{
  \institution{The University of Tokyo}
  \city{Tokyo}
  \country{Japan}
}

\author{Chihiro Shoda}
\email{chihiro.shoda@star.rcast.u-tokyo.ac.jp}
\orcid{0009-0008-4155-4217}
\affiliation{
  \institution{The University of Tokyo}
  \city{Tokyo}
  \country{Japan}
}

\author{Nanami Kotani}
\email{nanami.kotani@star.rcast.u-tokyo.ac.jp}
\orcid{0009-0005-6534-4589}
\affiliation{
  \institution{The University of Tokyo}
  \city{Tokyo}
  \country{Japan}
}
\renewcommand{\shortauthors}{Asahi, Sonoyama, Shoda, and Kotani}

\begin{abstract}
    Concentration is significant for maximizing potential in any activity. However, traditional methods to improve it often lack direct and natural feedback. We propose an innovative and inspiring VR system to experience and improve concentration. In the experience of pulling out the holy sword, which cannot be achieved simply by force, the player receives multimodal concentration feedback in visual, auditory, and haptic senses. We believe that this experience will help the user confront his/her concentration and improve the ability to control it consciously.
\end{abstract}


\begin{teaserfigure}
  \includegraphics[width=\textwidth]{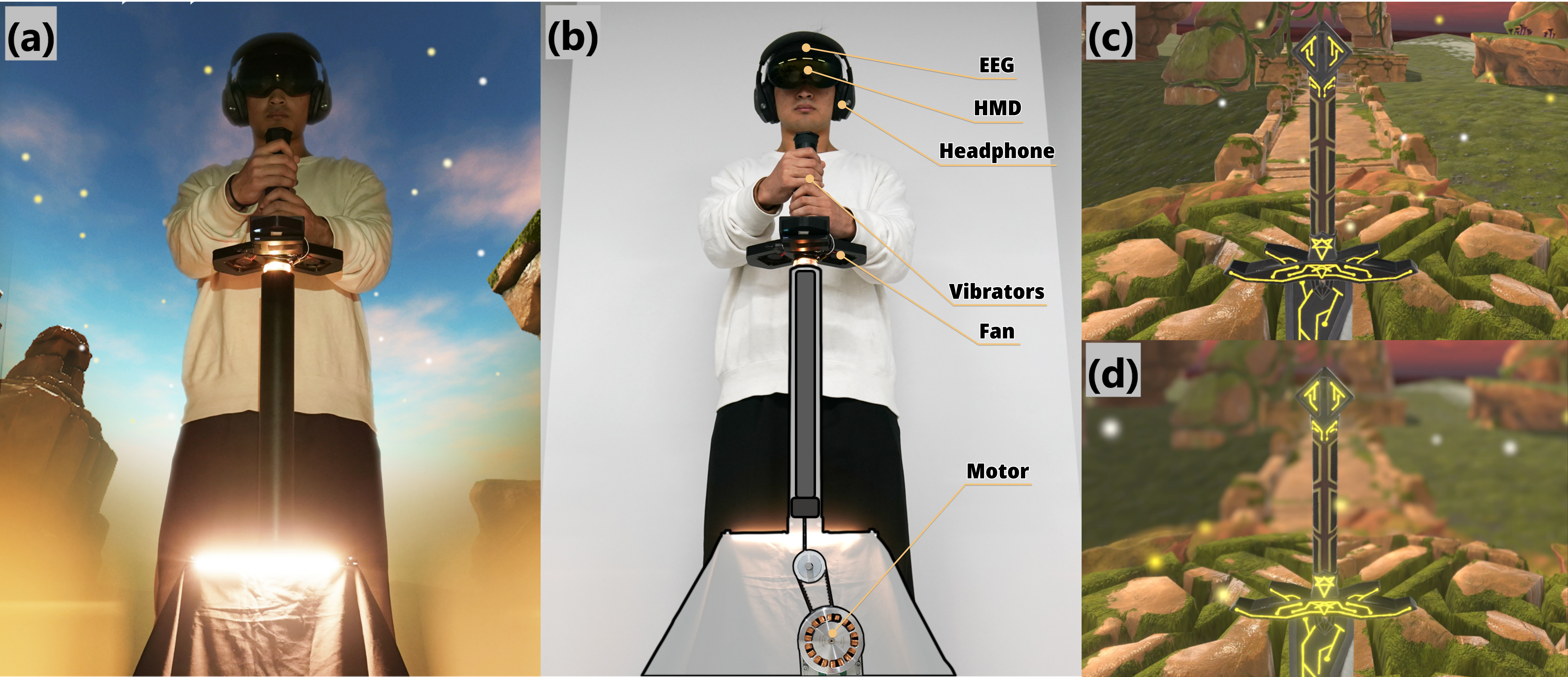}
  \caption{A player concentrates on pulling out the holy sword. (a) Appearance of the experience. (b) System design. (c) VR screen during lower states of concentration. (d) VR screen during higher states of concentration.}
  \Description{Overview of the proposed system}
  \label{fig:teaser}
\end{teaserfigure}


\maketitle

\section{Introduction}
Mindfulness, or the self-regulation of one's ability to concentrate nonjudgementlly on the present experience, is extremely important for maximizing one's potential in sports and other situations \cite{KEE2008393}. However, concentration does not directly affect the world, so it is difficult to enhance consciously. In recent years, biofeedback technology has been implemented to improve concentration using various wearable devices. One study attempts to combine Virtual Reality (VR) and Brain-Machine Interfaces (BMI) to achieve a higher concentration level by changing the VR space immersively during the concentration state\cite{inproceedings}. However, as with many attempts, real-time concentration feedback is based on a method that displays numerical values on a screen. Using this method, the user must once remove his/her eyes from the object of concentration and check the discrete change in numerical value on the screen when receiving feedback, which inevitably leads to a dispersion of concentration and interruption of the action. Therefore, receiving discrete feedback from a device outside the user's awareness during action is ineffective, and it is important to design a sequential feedback method that encourages concentration without obstruction.

\section{System design and implementation}
\begin{figure}[ht]
  \centering
  \includegraphics[width=\linewidth]{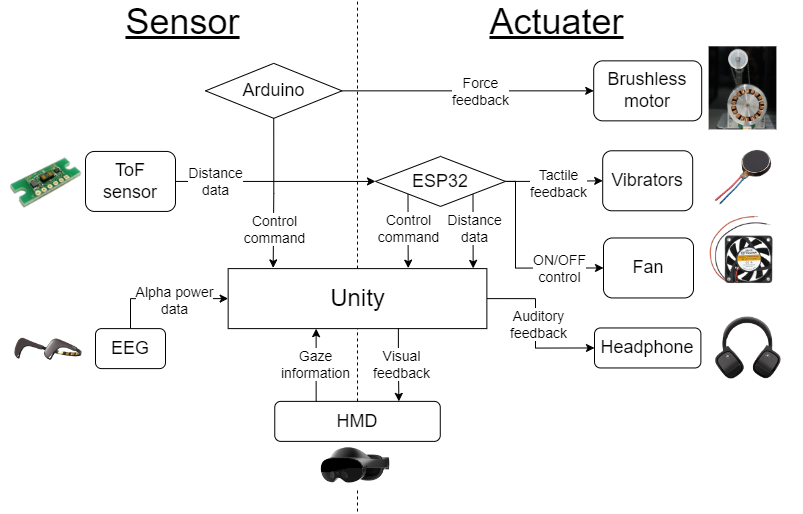}
  \caption{System architecture}
  \Description{Sensor and actuator}
  \label{fig:system}
\end{figure}

The scenario is a symbolic success story of pulling out a holy sword buried in an altar with concentrated effort.
The holy sword in VR moves in sync with the real-world hardware, which cannot be pulled out simply by force. However, increasing concentration through a feedback system combining three different modalities (visual, auditory, and haptic) makes the sword easier to pull out.
Therefore, the system can be mainly divided into concentration estimation and feedback parts. Other than that, to increase the efficacy of the successful experience, the system incorporates a mechanism that blows wind to the player when the sword is pulled out. The whole system architecture is shown in Fig.\ref{fig:system}.

\subsection{Measuring System}
In frequency bands within the Electroencephalography (EEG) power spectrum, alpha activity (8–13 Hz) plays an important role in attention by supporting processes within the attentional focus and blocking processes outside its focus. Therefore, alpha activity is said to reflect internal attention\cite{benedek2014alpha}.
In this project, we combine alpha waves from a frontal EEG with gaze information from a head-mounted display (HMD) to estimate concentration. Gaze information is used to suppress changes in spatial attention because it causes alpha modulation\cite{KERR201196}.

\subsection{Feedback System}
For the feedback system, we implemented visual feedback through the HMD, auditory feedback through headphones, and haptic feedback using vibrators and a motor based on the real-time estimated state of concentration.
Feedbacks are designed to reduce external stimuli sequentially, thereby inducing internal concentration.
The player's visual, auditory, and somatosensory attention is naturally spatio-temporally centralized as the player concentrates so that internal attention increases. 
Visually, the VR screen changes as the field of vision narrows and the surroundings slow down. 
Auditory, the surrounding environmental sound from the headphones becomes smaller and slower. 
Haptically, tactile sensations to the player's entire palm through the multiple vibrators in the sword grip and the traction force from the motor decrease.

\section{Game scenario}
During the experience, the player becomes the brave who is destined to pull out the holy sword. Before pulling out the sword, we calibrate by measuring the concentration and default states to obtain the relative value of concentration within an individual. While pulling out the sword, the player will be asked to apply a certain amount of force upward on the sword. The sword will be naturally pulled out if the player can enhance concentration through feedback.
In order to avoid applying too much force, the system measures the speed at which the sword is pulled out to ensure that the player pulls out slowly. In addition, Monsters' screams and other disturbing sounds can be heard while pulling out the sword, but the player must continue to concentrate and achieve in time.

\section{Discussion and future work}
This study is the first to combine visual, auditory, and haptic feedback in concentration feedback. We believe it is helpful to use non-visual modalities for concentration, which is mainly dependent on vision, and we will conduct experiments to determine which modality is appropriate for feedback. In addition, we need to clarify whether a linear mapping of the estimated value or a time-series average value is more appropriate for real-time awareness of concentration. Moreover, verifying the concentration estimation system's reliability through user tests will improve the quality of the experience. There is room for further study on how to achieve this system's goal of helping the user confront his/her concentration and improve the ability to control it consciously.

\bibliographystyle{ACM-Reference-Format}
\bibliography{reference}

\end{document}